\documentclass[preprint,aps,double]{article}

\usepackage{graphicx}
\usepackage{subfigure}
\usepackage{mwe}
\begin{document}

\title{
Orientational properties of the HGO system in a slit geometry in two-dimensional and three-dimensional case
from Monte Carlo simulations and Onsager theory revisited. }

\author{Agnieszka Chrzanowska}
\vspace{6pt}
Department of Physics, Krak\'{o}w University of Technology,
ul.\ Podchor\c{a}\.{z}ych 1, 30-084 Krak\'{o}w, Poland.
e-mail: a.chrzanowska@pk.edu.pl

\maketitle

\begin{abstract}
A problem of the orientational and density structure properties of a confined three-dimensional (3D) and two-dimensional (2D) Hard Gaussian Overlap (HGO) ellipsoids has been revisited
using the Onsager-type  second virial approximation of Density Functional Theory (DFT) and constant-pressure Monte-Carlo (MC)
simulations.
At the walls the assumed particles in 3D are forced to exhibit planar alignment.  In the nematic as well as in the smectic regime  particles situated apart from the walls attain homeotropic arrangement. This unusual bistable rearrangement is named as the eigenvalue exchange problem of the order parameter tensor. At the same time a bistable arrangement is not observed in the two-dimensional case of the same system. Comparison of the DFT theory and MC  simulation results has been given.
Whereas comparison of the orientational properties obtained from MC simulations and DFT theory is reasonable for a large range of densities, it does not concern
the density profiles. In denser systems differences become larger.
It occurred, however, that by manipulating  degree of  penetrability of the particles at the walls one can influence the surfacial density which improves  comparison.
 A discussion upon the problem what factors promote simultaneous existence of planar and homeotropic arrangement in a confinement has been provided.

\end{abstract}

\vspace {0.5cm}

\section{Introduction}
Due to application issues
interactions with surfaces is one of the most important factors that liquid crystalline community is interested in.
In the liquid crystalline cells  used widely in technological applications molecules are anchored at surfaces and the state of the system
in the cell is the result of the interplay of this anchoring influence with the electric field interaction with the whole liquid crystal (LC) system.
Whereas tailoring this anchoring in practical applications has been already mastered, knowledge on its
molecular origin is still incomplete
and many attempts have been undertaken so far to understand
physics of this phenomenon and its influence on the orientational properties of the system.
It was already known  more than 30 years ago \cite{SluckinPoniewier,Jerome,TelodaGama} that
in the presence of substrates mean local densities as well as orientational properties become inhomogeneous, mostly within surfacial areas. If the walls separations are of a few molecules lengths this modulations may influence also middles of the samples.
So far extensive studies on confined LCs comprise purely computer simulations
\cite{Barmes,BarmesHardNeedle,Cheung,Rene,Marjolein,Palermo,Lange2002Comp,Lange2002JChemPhys,
Greschek2010,Schoen,Lange2002Eur}, as well as
available theoretical approaches \cite{Sluckin1,Sluckin2,TeixeiraGayBerne,Cheung2004, Malijewski2010,Moradi,Avazpour} or both of them applied and compared for  the same system \cite{Allen,Chrzanowska2001,Cleaver2001,Deck,TeixeiraBarmesDeck,Teixeira2004}.

The most elaborated systems are ultra thin samples where the walls separations are less the $10$ molecules lengths
and the anchoring conditions are assumed as  symmetric (planar or homeotropic) or hybrid (planar on one side and homeotropic on the other side of the sample) conditions at the walls.
Theoretical approaches of such ultra thin systems started from the Landau--de Gennes descriptions \cite{Sluckin1, Sluckin2}, yet, as the authors noticed,  they suffered from many approximations.
More fruitful theories seem to be the ones of the
microscopic origin, like the density functional theory or the Onsager theory in the case of hard bodies.
Two types of rod like  particles shape are usually studied: of  ellipsoidal \cite{Avazpour,Chrzanowska2001,Cleaver2001,Deck,TeixeiraBarmesDeck,TeixeiraBarmes,Teix2016} or of cylindrical shape \cite{Rene, Marjolein,Malijewski2010,Heras2005,Heras2006}.

One of the most popular hard core ellipsoidal particles are the  objects defined by the Hard Gaussian Overlap rule \cite{Velasco,Padilla, Miguel} that stems out of the Gay-Berne potential \cite{GayBerne}.
To consider LCs in confinement
this interaction must be supplemented with the wall-particle interactions.
The simplest interaction  with the walls that is used is
the simple hard
needle-wall (HNW) surface potential.
This potential assumes that the interaction is governed by the behavior of a hard needle which goes throughout the particle
along its long axis \cite{BarmesHardNeedle,Moradi,Chrzanowska2001,Teixeira2004} and interacts with a flat hard surface. This idea is very convenient to use in the studies also  because of the fact that by changing the length of such a needle it is possible to moderate the degree of
of substrate penetrability.

Another idea to model surface interactions is to use a contact function that describes the walls as built from spheres or other HGO particles or to use special functions \cite{Avazpour}.
In \cite{Barmes}, for instance,  a special contact function has been developed
to characterize a surface potential capable of
exhibiting both homeotropic and planar anchoring alignments which promote appropriate alignment throughout the whole sample.
This surface potential allowed to observe the so-called
 bistable anchoring - the existence of the planar and homeotropic alignment throughout the whole sample depending upon the parameters used in  the contact function.
 Application of the above  idea  has been further continued in  \cite{Avazpour}.
 The word "bistable" is used here to show  possibility to obtain different arrangements
 with the use of the same surface contact function but with different parameters.
 This theoretical attempt  is aimed at describing
 the display cells that possess two optically distinct surface stabilized arrangements.
 It should be noted that the same word can be, however, used in different context --
when two types of arrangement, planar and homeotropic, exist at the same time in one sample.

In what follows we will focus on the effect found in the confined HGO system where the particles at the walls are planar and
the particles within the sample are perpendicular to the walls.
Section \ref{Model} presents the model, Section \ref{Onsager} introduces basic formulas of  the Onsager DFT approach. Section \ref{3D} and
Section \ref{2Dhgo} provide the results for three- and two-dimensional cases, respectively.
Finally Section \ref{discu} provides a discussion on the results.

\section{Model}
\label{Model}
We consider ellipsoidal hard  particles that interact through the potential
\begin{eqnarray}
U_{ij}({\bf{a}}_i,{\bf{a}}_j,{\bf{R}}_{ij} ) =
\left\{
\begin{array}{rllrllr}
0 & \;{\rm{if}} & r_{ij} \geq \sigma ({\bf{a}}_i,{\bf{a}}_j,{\bf{r}}_{ij} ) \\
\infty &\;{\rm{if}} & r_{ij} < \sigma ({\bf{a}}_i,{\bf{a}}_j,{\bf{r}}_{ij} ), \\
\end{array}\right.
\label{pot}
\end{eqnarray}
where ${\bf{a}}_i$ is the unit vector pointed along the particle $i$ and describing its orientation and ${\bf{R}}_{ij} $ is the vector connecting centers of the particle $i$ and $j$.
Vector ${\bf{r}}_{ij}$ is the unit vector along the direction  ${\bf{R}}_{ij} $.
$\sigma ({\bf{a}}_i,{\bf{a}}_j,{\bf{r}}_{ij} )$ is the particle shape function used in the Gay-Berne potential \cite{GayBerne} which  serves here as the contact function.  Its form follows
\begin{equation}
\sigma ({\bf{a}}_i,{\bf{a}}_j,{\bf{r}}_{ij} ) = \sigma_0 \left(
1-\frac{1}{2}\chi \left[
\frac{({\bf{r}}_{ij} \cdot{\bf{a}}_{i}+{\bf{r}}_{ij} \cdot{\bf{a}}_{j})^2 }
{1+\chi ({\bf{a}}_{i} \cdot{\bf{a}}_{j})}
+
\frac{({\bf{r}}_{ij} \cdot{\bf{a}}_{i}-{\bf{r}}_{ij} \cdot{\bf{a}}_{j})^2 }
{1-\chi ({\bf{a}}_{i} \cdot{\bf{a}}_{j})}
\right]
\right)^{-1/2}
\end{equation}
With respect to $X,Y$ and $Z$ coordinates this function is of the square type hence
the condition (\ref{pot}) describes  the surface of an ellipsoid in 3D and an ellipse in 2D.
Anisotropy of the ellipsoid is provided by the factor $\chi=\frac{\kappa^2-1}{\kappa^2+1}$, where $\kappa$ is the length to breadth ratio of the particle, $\kappa=L/\sigma_0$.
The above formulas are very convenient to use in the Monte Carlo simulation as well as in the Onsager theory.
To study arrangement of the particles confined between substrates one needs also to
impose conditions forcing orientational arrangement directly at the walls.
For homeotropic orientational arrangement it is sufficient to forbid the particles centers
to go beyond the walls. To induce planar arrangement the following condition has been proposed

\begin{eqnarray}
U_i(\theta_i,y_i)=
\left\{
\begin{array}{rllrllr}
0 & \;{\rm{if}} & \;\;|z_i-z_{wall}| \geq \lambda {\rm{cos}}(\theta_i), \\
\infty &\;{\rm{if}} & \;\; |z_i-z_{wall}| > \lambda {\rm{cos}}(\theta_i), \\
\end{array}
\right.
\label{potWall}
\end{eqnarray}
where $\lambda$ is the parameter responsible for embedment of the particle within the surface.
For $\lambda = \frac{L}{2}$ the condition (\ref{potWall}) is equivalent to the interaction of a hard needle of length $L$ with a hard surface.
As it was shown, for instance  in \cite{Teixeira2004}, by changing  $\lambda$ one can influence indirectly the density at the surface.

\section{The Onsager theory}
\label{Onsager}

The  Helmholtz free energy density functional in the second virial approximation  is given by
\begin{eqnarray}
& \beta F_{\rm{Helm}}[\rho ( {\bf{r}},{\bf{a}} ) ]=\int \rho({\bf{r}}_1,{\bf{a}}_1) \; {\rm log}  [ \rho({\bf{r}}_1,{\bf{a}}_1) -1] \;d{\bf{r}_1}d{\bf{a}}_1 & \nonumber \\
& -\frac{1}{2} \int \left(  {\rm exp}[-\beta U_{12}]-1
\right)\; \rho({\bf{r}}_1,{\bf{a}}_1)\; \rho({\bf{r}_2},{\bf{a}}_2)\; d {\bf{r}_1} d{\bf{a}}_1 d{\bf{r}_2} d{\bf{a}}_2  & \nonumber \\
&
-\mu \beta \int \rho({\bf{r}}_1,{\bf{a}}_1) \;d{\bf{r}_1}d{\bf{a}}_1,
&
\label{free_energy_functional}
\end{eqnarray}
where $U_{12}$ stands for the interaction potential,  $ ({\rm exp}[-\beta U_{12}]-1)$ is the Mayer function and
the reduced temperature is $\beta=\frac{1}{kT}$ with $k$ being the Boltzmann factor.
${\bf{a}}$ in the case of uniaxial particles is the vector along main molecular axis.
 $\rho$ provides probability of finding a particle at a given spatial position  ${\bf{r}}$ and oriented
according to the orientation of the vector ${\bf{a}}$.
$\mu$ is the chemical potential.
The probability function $\rho$ depends in 2D on the $X$ and $Y$ spatial coordinates that span the entire
surface and on one angular coordinate $\phi$, whose values belong to the interval $(0,2\pi)$.
In 3D $\rho$ depends on three coordinates, $X$, $Y$ and $Z$, and two angles $\theta$ (from the interval $(0,\pi)$) and $\phi$ (from the interval $(0,2\pi)$).
Routinely, $\rho$ is normalized to the number of particles $N$.
In case of interacting walls this formula must be supplemented with the term
describing interactions of particles with the walls.
\begin{eqnarray}
\beta F_{{\rm{walls}}}[ \rho ({\bf{r}},{\bf{a}}) ]= \beta \int
\rho ({\bf{r}},{\bf{a}})  \sum_i U^i_{{\rm{wall}}}\;\; d {\bf{r}} d{\bf{a}},
\label{wallInt}
 \end{eqnarray}
where $i$ is the number denoting  the wall.
These forms, (\ref{free_energy_functional}) and (\ref{wallInt}), have to be recalculated according to the system and particle geometry.
In the case of hard interactions the  Mayer function is equal to $-1$, when the particles overlap and $0$ otherwise, which, after integration over one set of spatial
variables,  results in the excluded volume $V_{{\rm{excl}}}$.

To obtain  the distribution function $\rho$ one has to minimize to whole system energy
with respect to the geometry of the outcome result, here $Z$ dependence and $\theta$ and $\phi$ in 3D (or $Y$ and $\theta$ in 2D).
\begin{eqnarray}
\frac{\delta (F_{{\rm{Helm}}}+F_{{\rm{walls}}} )} {\delta \rho}=0.
\end{eqnarray}
 This leads to the self-consistency equation, for 3D for instance as
 \begin{eqnarray}
{\rm log} \rho ({\bf{a}}_1,z_1)= -d  \int
V_{{\rm{excl}}} \rho ({\bf{a}}_2,z_2) dz_2 d{\bf{a}}_2
\label{self}
\end{eqnarray}
where $d=N/V$ stands for the averaged density. Note that in the case of inhomogeneous systems the local density is not constant and the profiles of such local densities is the subject of the study whereas the averaged density plays the role of the parameter in the theory.

The equation (\ref{self}) has to be solved now numerically
in a self-consistency manner.

Numerical scheme chosen for performing integrals is based on the Gaussian quadratures applied at the Gaussian points.
 The appropriate values, the excluded volume and distribution functions,  are stored then in the form of a matrix and, subsequently, used to find equilibrium solutions to the problem.
The use of the Gaussian quadratures
allows for diminishing the number of the integral function evaluations, which is especially crucial in the case
of multidimensional integrations inherent to the DFT theories of liquid crystals.
Due to the hindrance imposed by the walls, integrations over the polar $\theta$ angle cannot span the whole interval $(0-\pi)$ within the surface region.
The effective interval will depend here on the distance of the particle from the wall.
Due to the fact that we use matrix method in the calculations, where  values of $\rho$ are stored at the Gaussian points $z_i$,
we can also utilize  the angular Gaussian points within the interval $(\theta_{min}(z_i)-\theta_{max}(z_i))$ parametrized by $z_i$ values. In such a manner one can use always the whole set of the Gaussian points.
The numerical method with such $z_i$ adjusted Gaussian points
has been presented in \cite{ChrzanowskaMetoda}.
Note also that the Gaussian method is also convenient
to resolve problems of liquid crystals
with higher symmetry than nematics
\cite{AgnesferroMetoda1,AgnesSmekMetoda2}.

\section{Results for $3D$ HGO system.}
\label{3D}
In what follows we present properties of the above presented HGO system of particles.
Thickness of the particles is assumed here as $\sigma_0=1$ and the shape aspect ratio   as $\kappa=L/\sigma_0=5$.
Molecules  are placed between hard walls separated by their $4$ lengths.
In such an ultra thin sample the walls significantly influence the state of the whole system.
Additionally particles at the walls are allowed to immerse slightly in the walls as to
tailor  density at the surfaces which  may  decide, for instance,  on the number of density peaks in the smectic regime.

\begin{figure}[htp]
\begin{center}
\includegraphics[width=0.95\columnwidth]{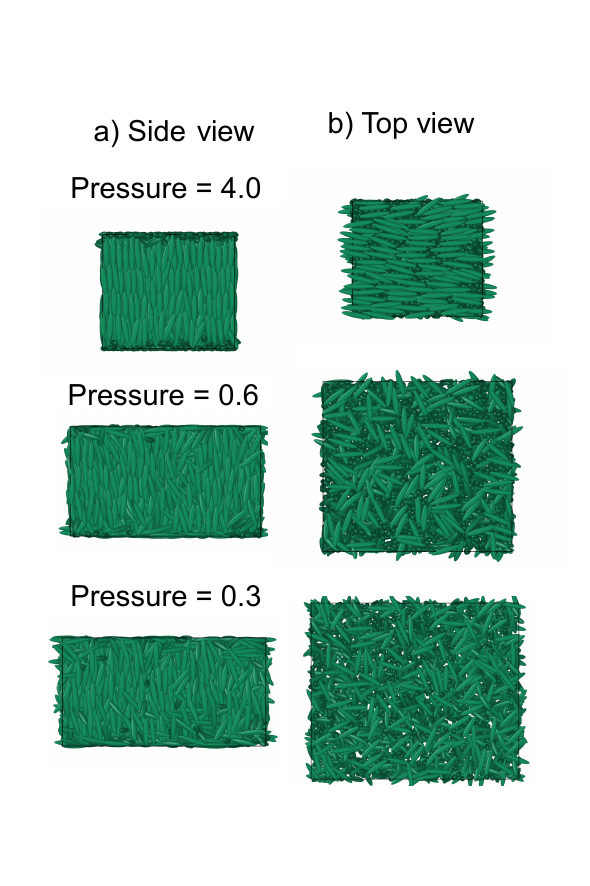}
\end{center}
\caption{Snapshots of the system from MC simulations. a)  side view (on the left) and b) top view (on the right).
For pressure equal to $0.3$ a nematic phase with disorder at the surfaces is seen, for $0.6$ an onset of a smectic like phase with disorder at the surfaces and for $4.0$ a well ordered smectic with 2D nematic order at the surfaces is presented.
}
\label{fig_elipki}
\end{figure}

\begin{figure}[htp]
\begin{center}
\vspace{0.8cm}
\includegraphics[width=0.85\columnwidth]{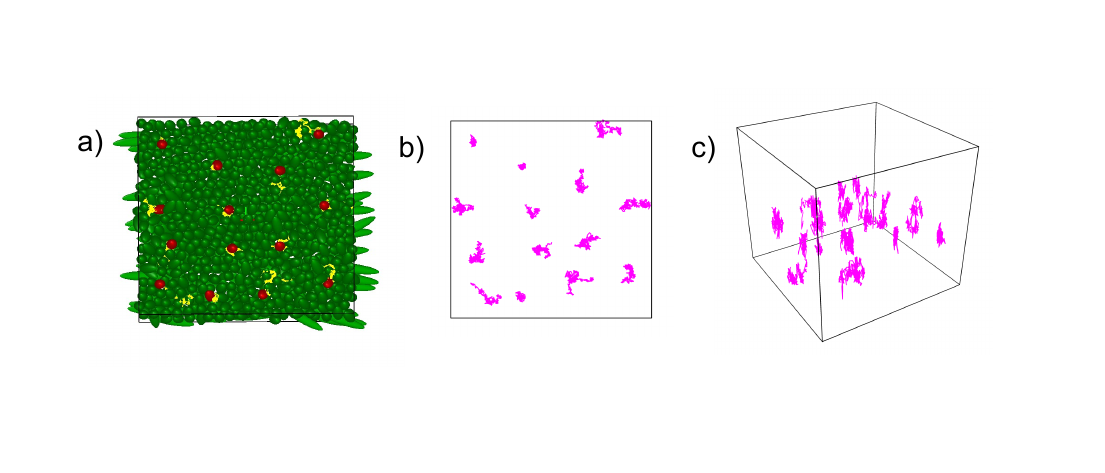}
\end{center}
\caption{Monte Carlo trajectories (pink lines) for a few chosen particles (given in a) in red color).
a) presents the top view at the layer from which the particles have been chosen. Ellipsoids are drawn in green colour. b) shows  only trajectories without viewing the particles, seen from the top, similarly as in a).
c) trajectories seen from a perspective showing movement of the particles between layers.
}
\label{fig_traje}
\end{figure}

In Figure (\ref{fig_elipki}) three different liquid crystalline states of the investigated three-dimensional system are presented from the Monte Carlo simulations. For the low pressure equal to $0.3$ a typical nematic arrangement is seen, where the particles inside the sample are ordered perpendicularly to the walls and these ellipsoids, which are at the walls, are planarly and chaotically arranged.
This is already worth paying attention  since an initial arrangement of the particles was parallel to the walls. During MC simulation the molecules inside the sample must have reoriented about 90 degrees.
For the  pressure equal to $0.6$  an onset of a smectic layering with orientational disorder of the particles at surfaces is observed. Again one sees here coexistence of planar and homeotropic arrangement.
For the  pressure equal to $4.0$
a well orientational order of the particles, which are arranged in layers, is seen.  At the surfaces one observes also
 a two-dimensional orientational order.
 By examining the Monte Carlo mobility of some chosen  particles of the system at the pressure $4.0$ in the form of MC trajectories,
 Figure (\ref{fig_traje}),
 one concludes that the system is of the smectic type, although at the first glance one may think of it as a solid.
Figure (\ref{fig_traje}a) shows the part of the system cut from the middle of the sample, so we see ends of the green ellipsoids.
Ellipsoids for which trajectories have been calculated are given in red. Trajectories themselves here are yellow.
From the panel (\ref{fig_traje}b), where only trajectories in the top view geometry are given, one observes that the particles can move from
site to site, hence they are not blocked in crystalline nodes. Perspective view at these trajectories in (\ref{fig_traje}c)  reveals also that the ellipsoids can relatively easily move from one layer to another one. This convince us that the system exhibits smectic ordering.

\begin{figure}[htp]
\vspace{0.8cm}
\includegraphics[width=0.45\columnwidth]{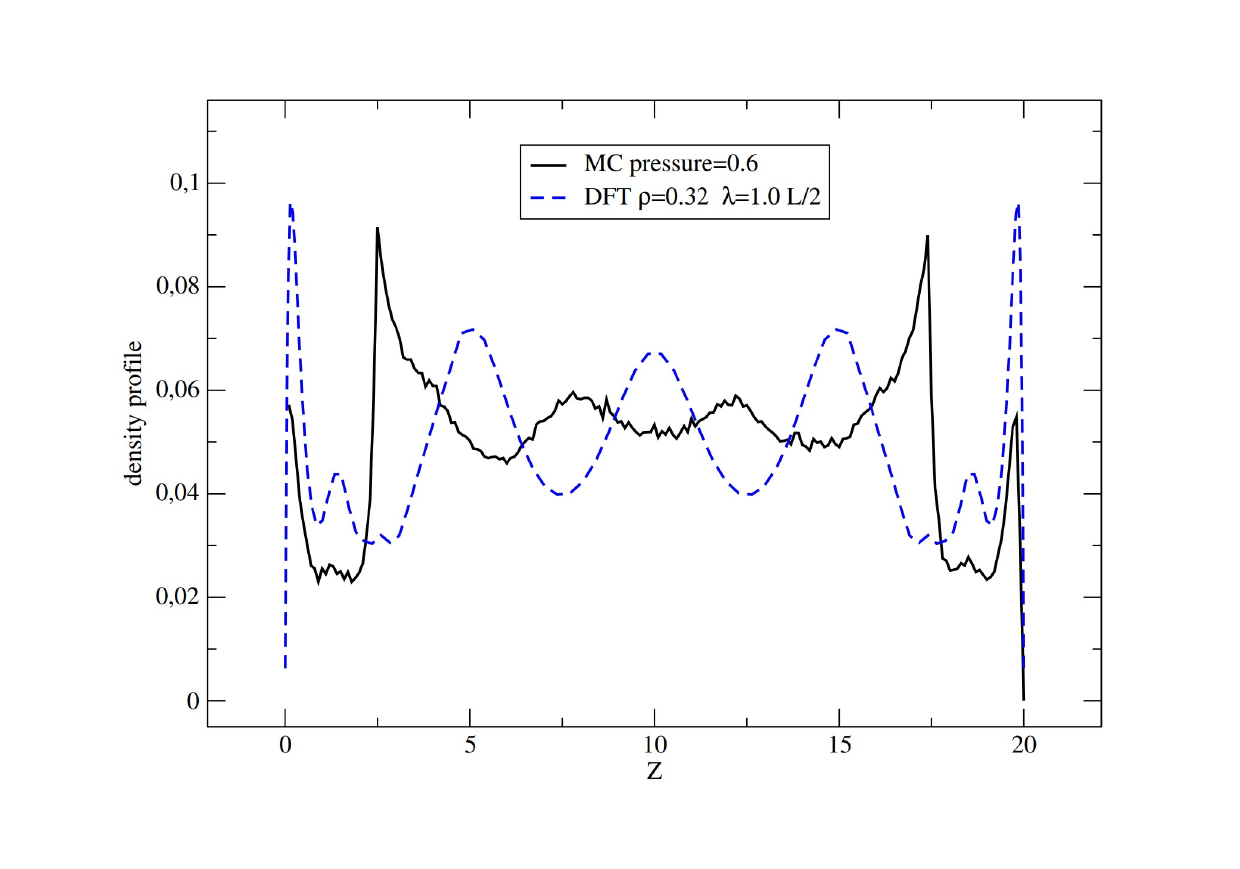}
\hspace{0.05\columnwidth}
\includegraphics[width=0.45\columnwidth]{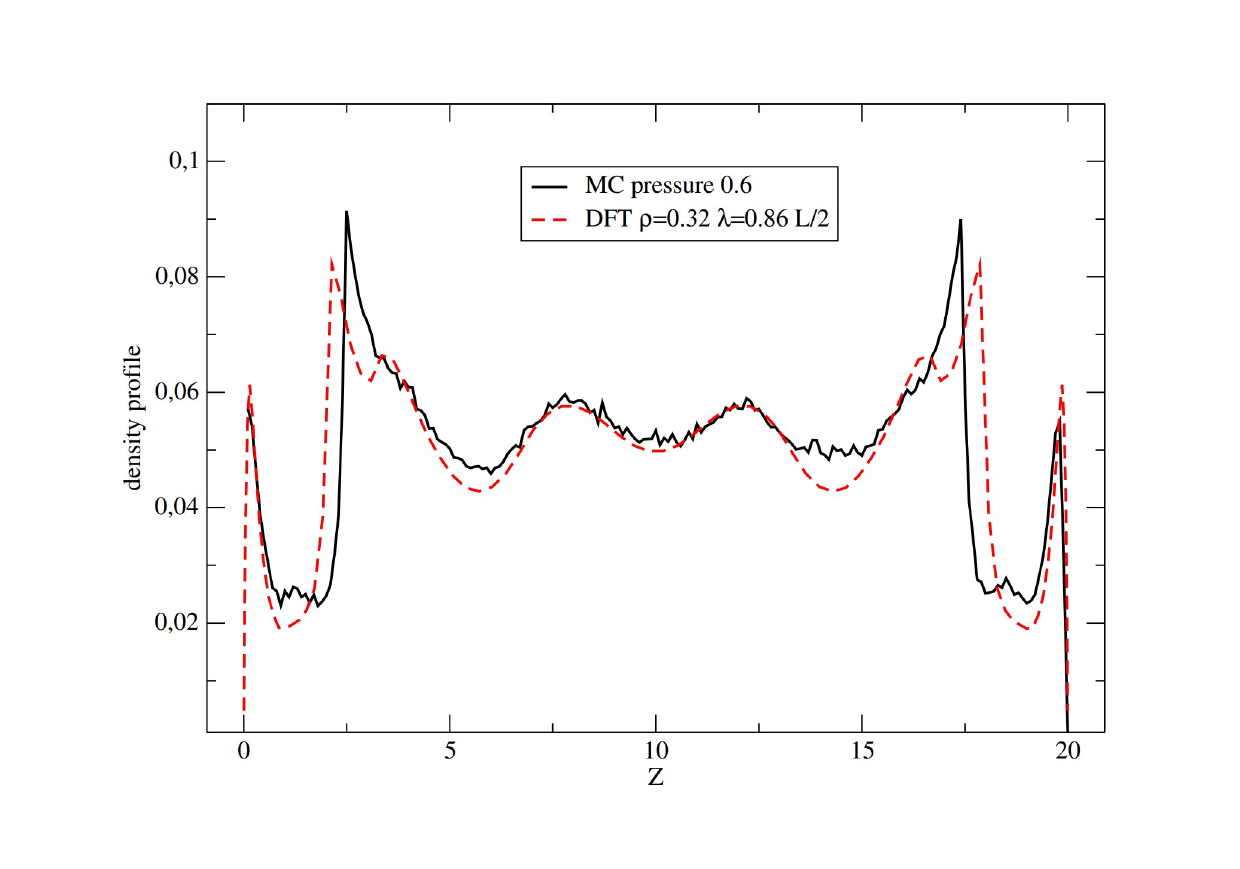}
\caption{Density profiles vs perpendicular to the walls distance $z$ for the cases with the penetrability parameter $\lambda=L/2$ (left) and $\lambda=0.86 L/2$ (right), for averaged density $0.32$ and pressure $0.6$.}
\label{fig_plan32}
\end{figure}

In Figure (\ref{fig_plan32}) an exemplary  density profile for a system has been provided for $d=0.32$ .
 A chosen value of pressure in the MC simulations is here $0.6$. The  average density $d$ used
in the DFT calculation has been chosen in such a way that the resultant density profile
seemed to be the closest match to the MC result with respect to the heights of the peaks.
We  do  not use at this point  any scalings like, for instance, Parsons scaling \cite{Parsons}, since it is not known to what extent it influences the surfacial regimes or whether it acts equally well in the bulk and in the surfacial areas. Additionally, we bear in mind that  the arrangement is already of the smectic type (the onset).
The aim of our procedure was simply to find a match from the DFT calculations to the MC results and too see how density at the walls influences the particles arrangement.

We see already, what previously was also highlighted in \cite{Teixeira2004}, that the second virial theory underestimates the numbers of layers. Here, DFT predicts strong peaks at the walls and three density peaks  (layers) in the middle of the sample.
In the MC simulations the peaks at the walls are much lower with very steep peaks  at the
distances from the walls where the particles  regain complete rotational freedom.
The number of peaks is  in the middle of the sample is two.
By changing the value of the penetrability parameter $\lambda$ in the DFT application one observes that much better comparison can be obtained.
 This can be seen in  Figure (\ref{fig_plan32}) on the right panel. Not only the number of peaks are correct but also their profiles are getting very close.

Orientational properties of the system are at best seen by studying the behaviour of the order parameter tensor defined as
\begin{equation}
{\bf Q}= \frac{1}{2}< 3 {\bf a} {\bf a}-{\bf{\delta}}  >.
\end{equation}
If the coordinate system is chosen to correspond to the symmetry of $\bf{Q}$ one needs only to look at the diagonal terms, since off diagonal terms are zero.
In our case $Z$ axis is perpendicular to the walls surfaces and $X$ and $Y$ axes are situated  within $XY$ plane.

\begin{figure}[htp]
\vspace{0.8cm}
\includegraphics[width=0.4\columnwidth]{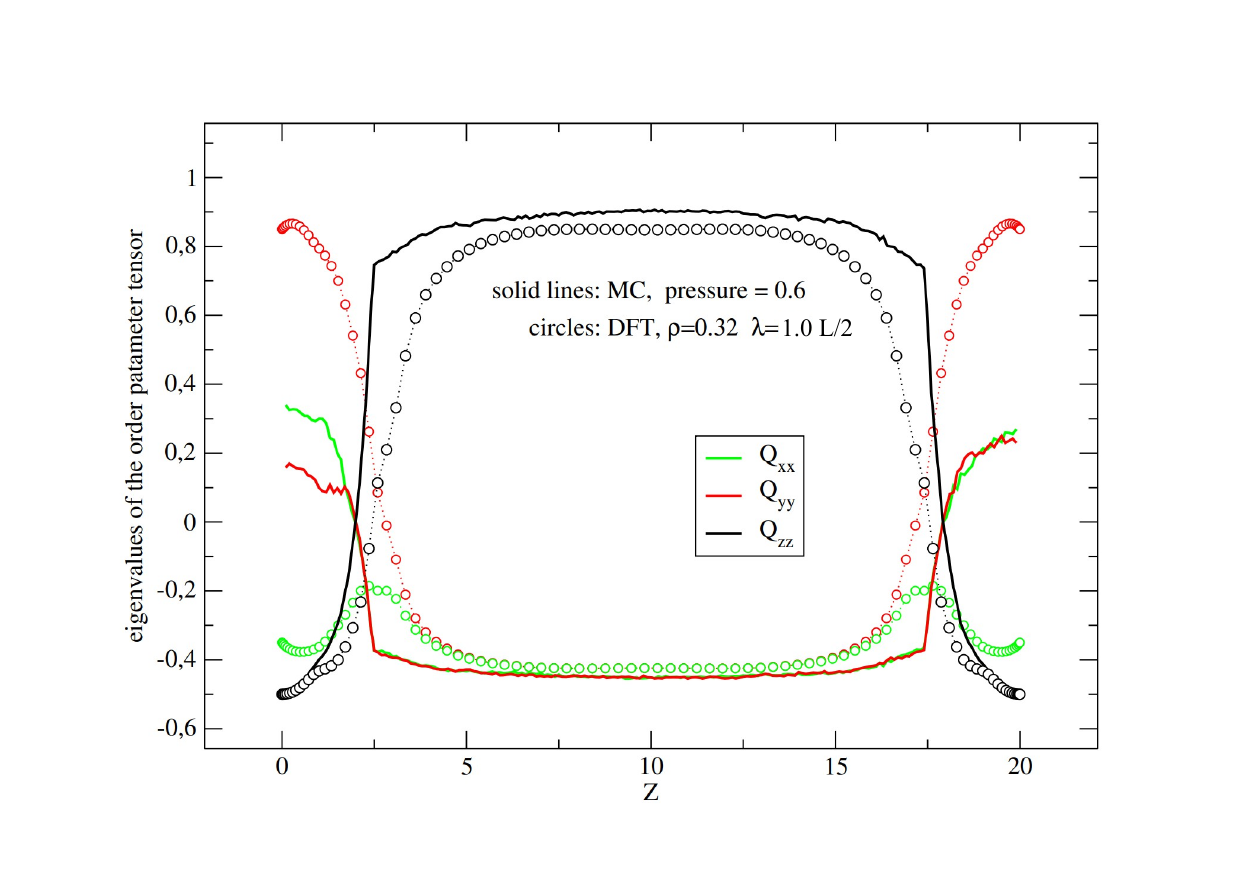}
\hspace{0.05\columnwidth}
\includegraphics[width=0.4\columnwidth]{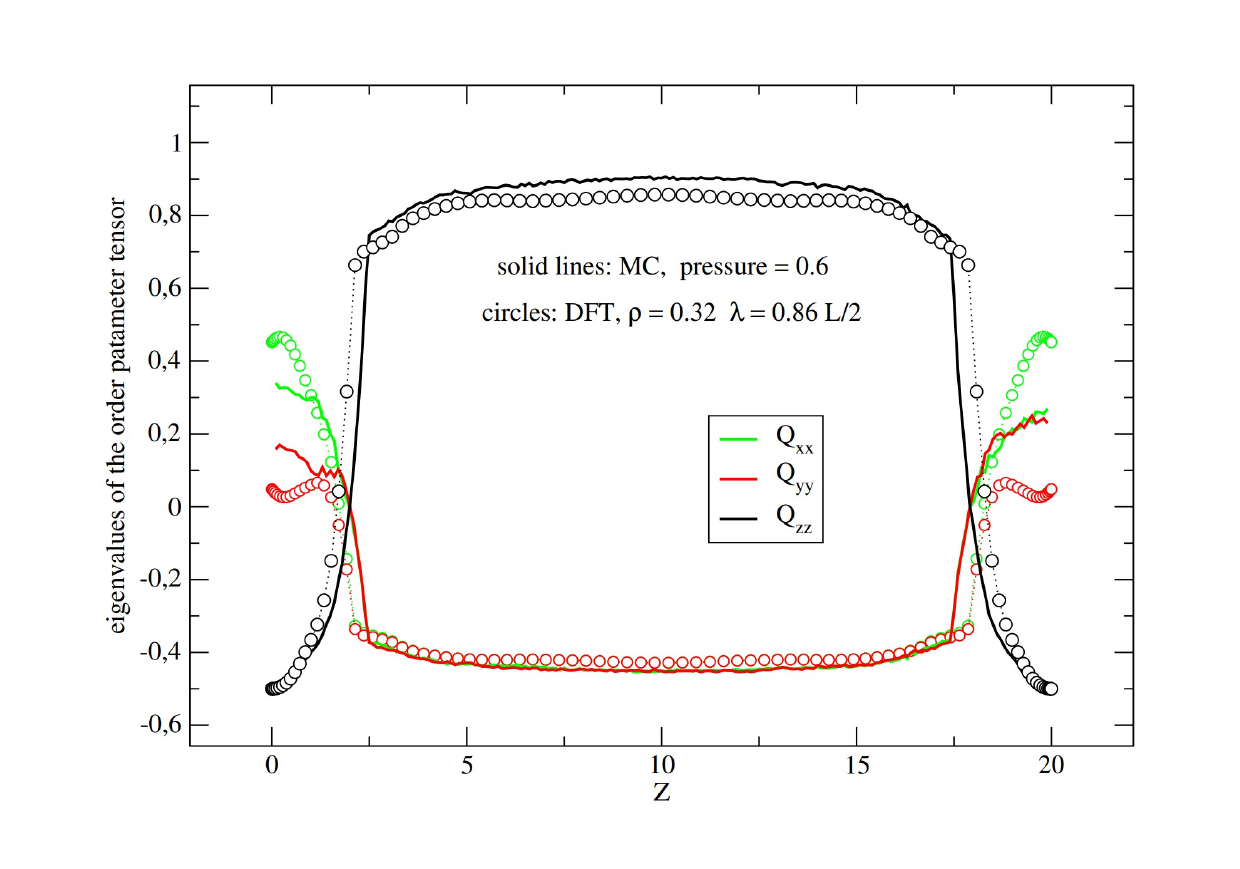}
\caption{Eigenvalues of the order parameter tensor for the cases with $\lambda=L/2$ (left) and $\lambda=0.86 L/2$ (right).}
\label{fig_Quplan32}
\end{figure}

Figure (\ref{fig_Quplan32}) shows eigenvalues of the order parameter tensor obtained from MC simulations for the case where pressure is equal to $0.6$  and their matching profiles obtained  from the DFT calculations.  Off diagonal components are not presented since they are at the level of zero (with some fluctuations in the MC results).
Two different penetrability parameters  have been here used. The meaning of these profiles is as follows. 

If the particles are positioned along $Z$ axis, $Q_{zz}$ is positive (for perfect alignment will be equal to 1). At the walls, when particles
are perpendicular to $Z$ axis, $Q_{zz}$ is equal to $-0.5$. Since $Tr {\bf Q}=0$ for uniaxial ordering two other eigenvalues must have opposite sign and be equal.
This  is seen in the middle of the sample
in both cases, without and with renormalization of the surface density caused by the change of $\lambda$ parameter.
In the left panel  Figure (\ref{fig_Quplan32}) the DFT theory predicts biaxial ordering close to the walls. All three eigenvalues are different. This is so since particles are (almost) confined to the plane and within this plane they are also ordered in one direction.
For the pressure equal  $0.6$
in Figure (\ref{fig_elipki}) we do not see such an arrangement. Top view of the system shows chaotic arrangement  in the surface plane.
By changing penetrability parameter (the right panel of (\ref{fig_Quplan32})), which influence the density at the walls,
the orientational profiles become more close to the MC results. Because of the visible modulation of the density profile in the middle of the sample this case is of the smectic type (onset of this ordering).

\begin{figure}[htp]
\begin{center}
\vspace{0.8cm}
\includegraphics[width=0.55\columnwidth]{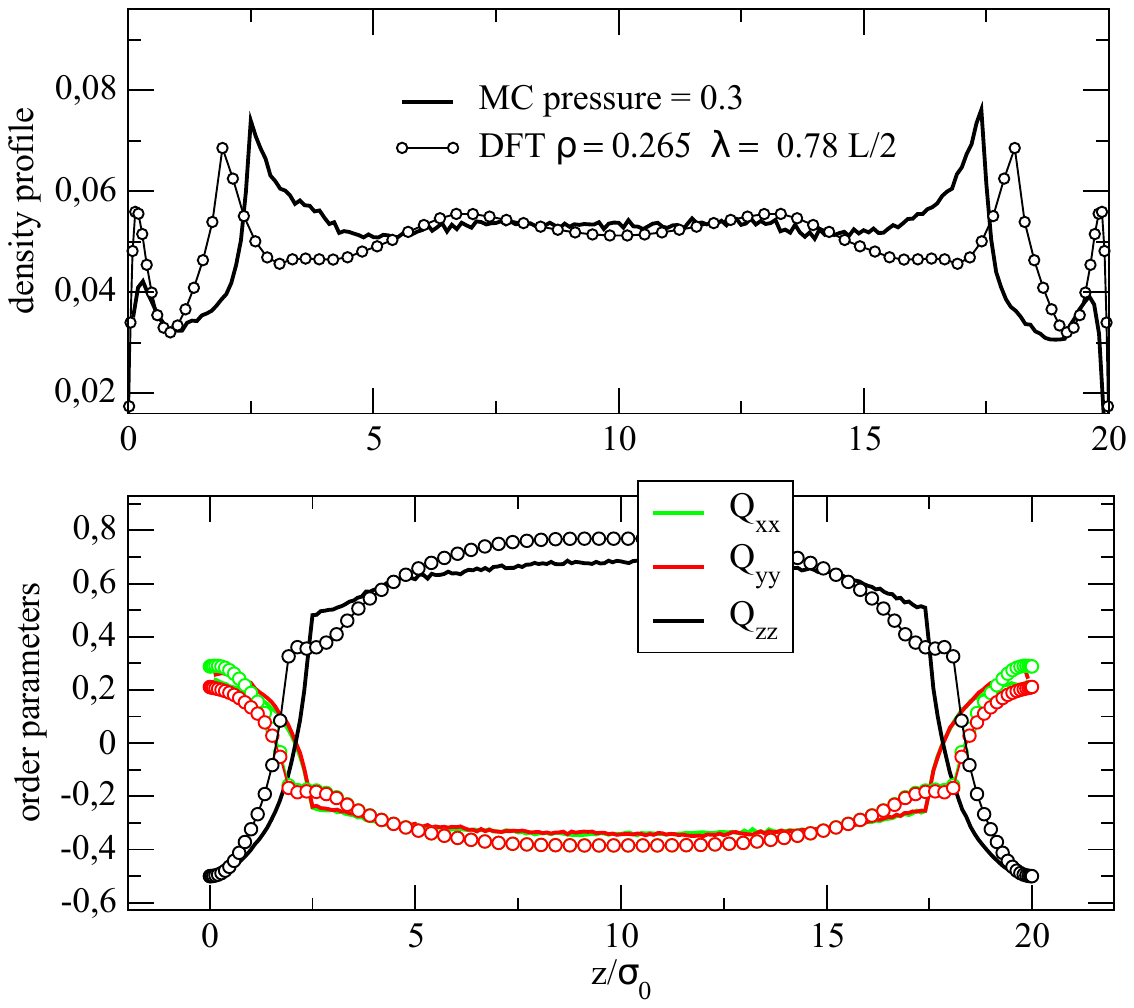}
\hspace{0.05\columnwidth}
\end{center}
\caption{Density profiles vs perpendicular to the walls distance $z/\sigma_0$ for the cases with the penetrability parameter $\lambda=L/2$ (left) and $\lambda=0.86 L/2$ (right) for nematic arrangement.}
\label{fig_press03}
\end{figure}

\begin{figure}[htp]
\begin{center}
\vspace{0.8cm}
\includegraphics[width=0.55\columnwidth]{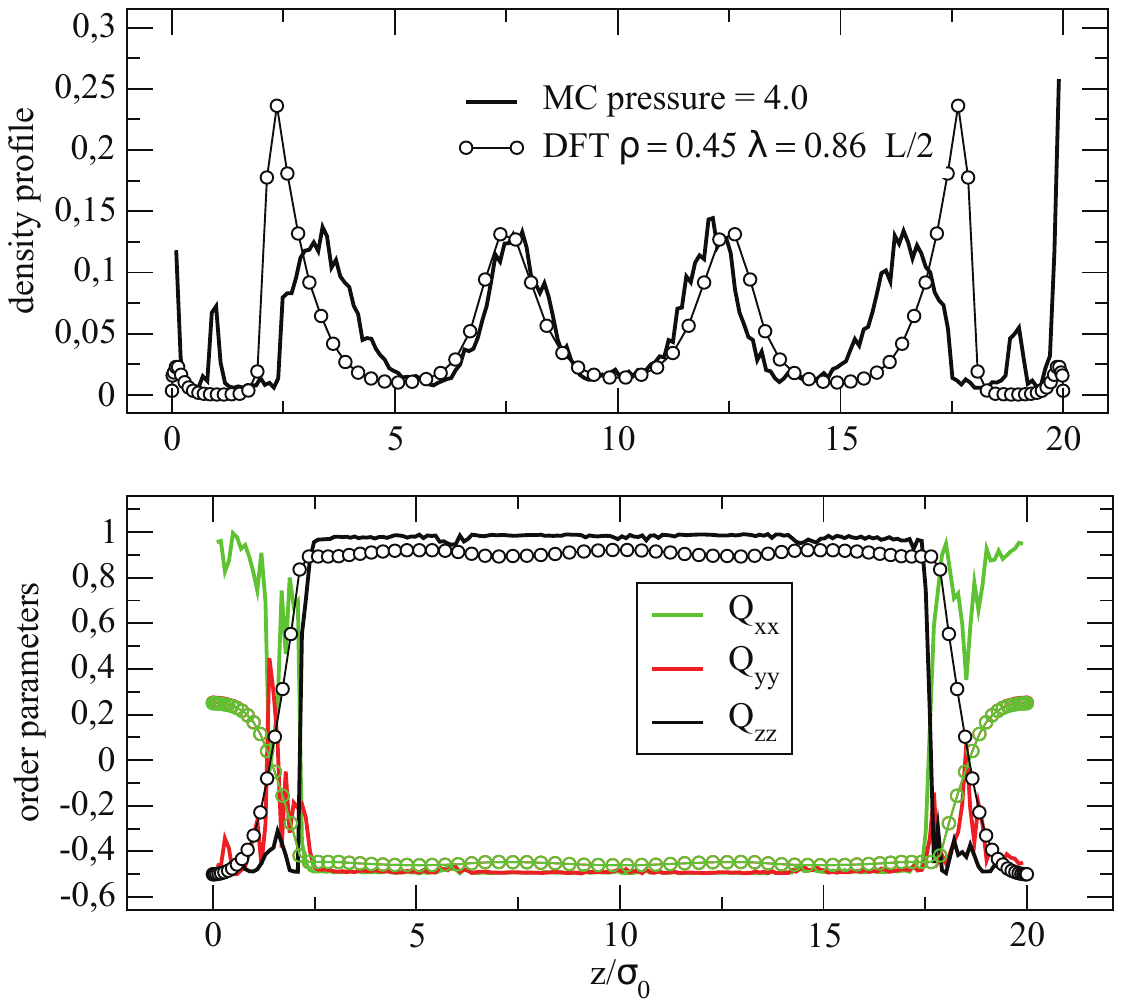}
\end{center}
\caption{Density profiles vs perpendicular to the walls distance $z/\sigma_0$ for the cases with the penetrability parameter $\lambda=L/2$ (left) and $\lambda=0.86 L/2$ (right) for smectic arrangement.}
\label{fig_press40}
\end{figure}

In Figure (\ref{fig_press03}) we present exemplary results for the nematic region.
Here again the profiles obtained from simulation and theory are quite similar. The height of the density peaks are almost  the same, yet due to the use of different $\lambda$'s the positions of them are different. On the other hand, the eigenvalues of ${\bf{Q}}$ seem very similar.
It also turns out that such  comparisons for denser and well ordered smectic  phases
(like in Figure (\ref{fig_press40}))
is also reasonable, although still some departures are seen. The density parameter in the DFT calculation was chosen as to obtain the similar heights of the peaks inside the sample. In this case however, the DFT results does not reproduce small second  peaks at the walls and other peaks from the DFT close to the walls are larger.

In all the above examples we observe a simultaneous bistable ordering.
In the next section we consider a two-dimensional counterpart of the current system
in order to see, whether such a bistability is also present.
\section{Results for $2D$ HGO system.}
\label{2Dhgo}
In this section we would like to check whether bistable simultaneous ordering will be also present in the two-dimensional case.

\begin{figure}[htp]
\begin{center}
\hspace{-1.2cm}
\includegraphics[width=0.45\columnwidth]{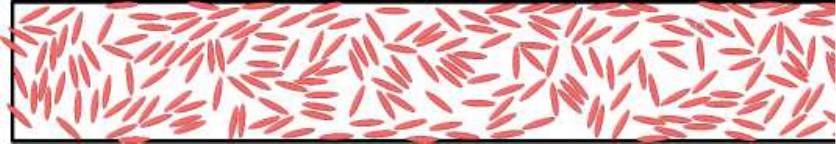}\\
\includegraphics[ width=0.65\columnwidth]{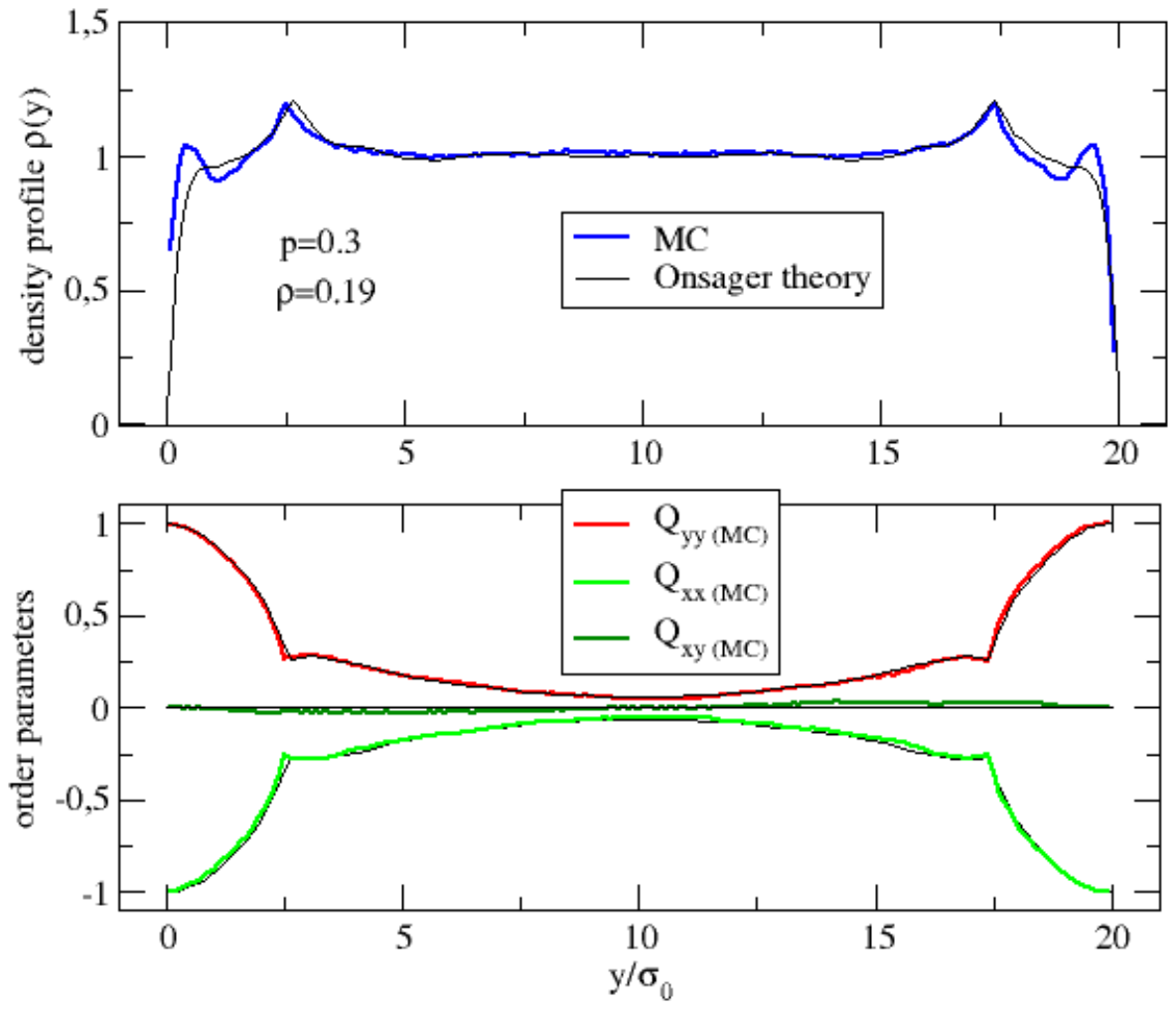}
\end{center}
\caption{Density profile and components of the order parameter tensor for a dilute system drawn in the direction perpendicular to the walls ($Y$ direction).
 Above:  a view at the part of  a dilute 2D HGO system with the walls placed at up and down levels. }
\label{2d_019MC}
\end{figure}

\begin{figure}[htp]
\begin{center}
\hspace{-1.2cm}
\includegraphics[width=0.35\columnwidth]{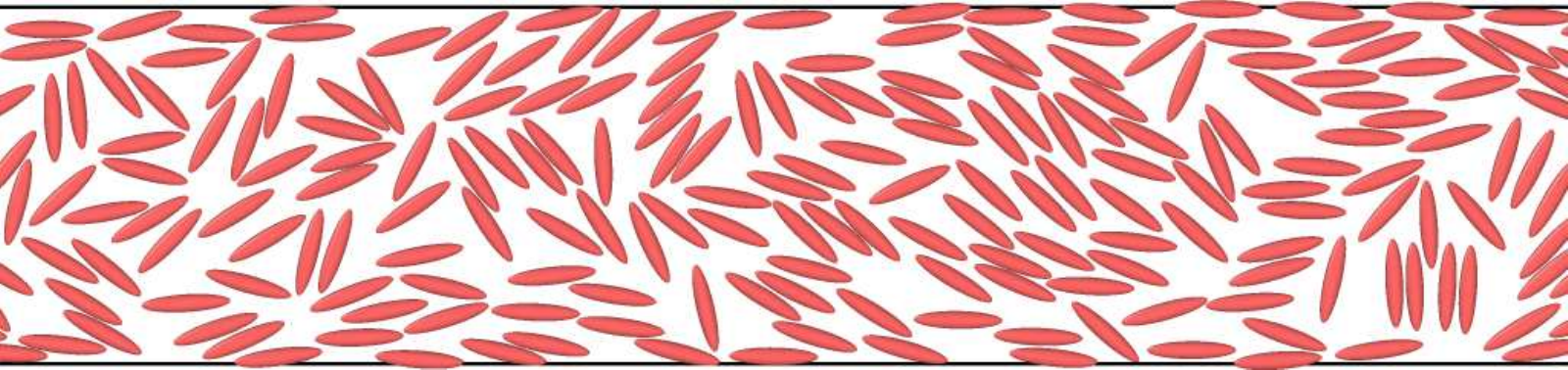}\\
\includegraphics[width=0.65\columnwidth]{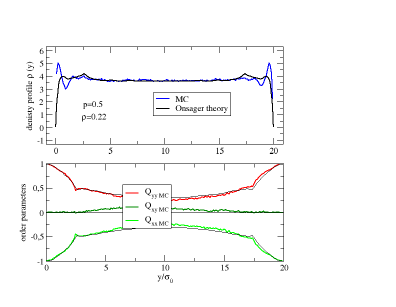}
\end{center}
\caption{Density profile and components of the order parameter tensor for the case
at the onset of the nematic phase. Above:  a view at the part of  a  2D HGO system with the walls placed at up and down levels. }
\label{2d_022MC}
\end{figure}

 In Figure (\ref{2d_019MC}) a low density case with isotropic phase in the middle of  the sample is presented. Both types of profiles exhibit a good comparison of the simulation and theoretical results. Comparison of the orientational profiles are very good, yet in the case of the density profiles there are departures at the walls. No bistable ordering is seen.
Similar effects are seen in a denser system with the onset of the nematic phase presented in Figure (\ref{2d_022MC}).

\begin{figure}[htp]
\vspace{0.8cm}
\begin{center}
\includegraphics[width=0.4\columnwidth]{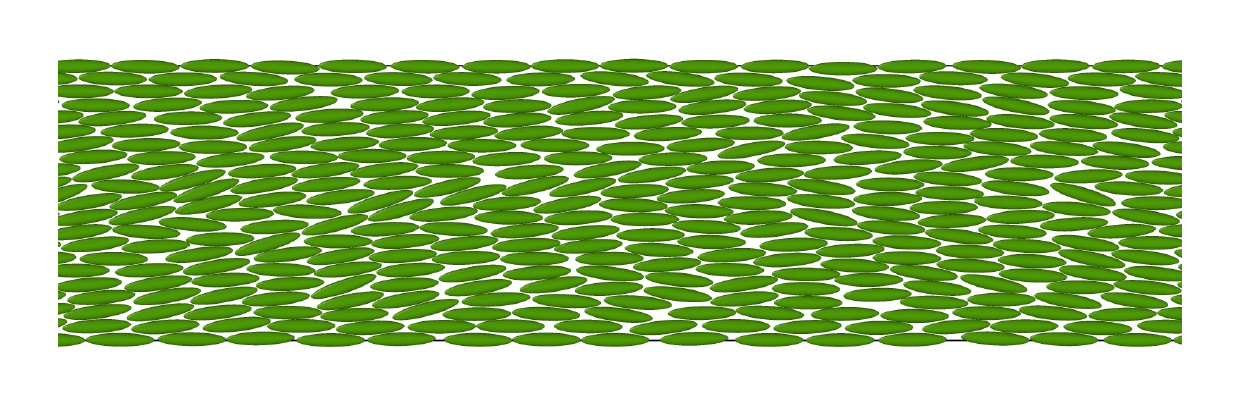}
\includegraphics[width=0.65\columnwidth]{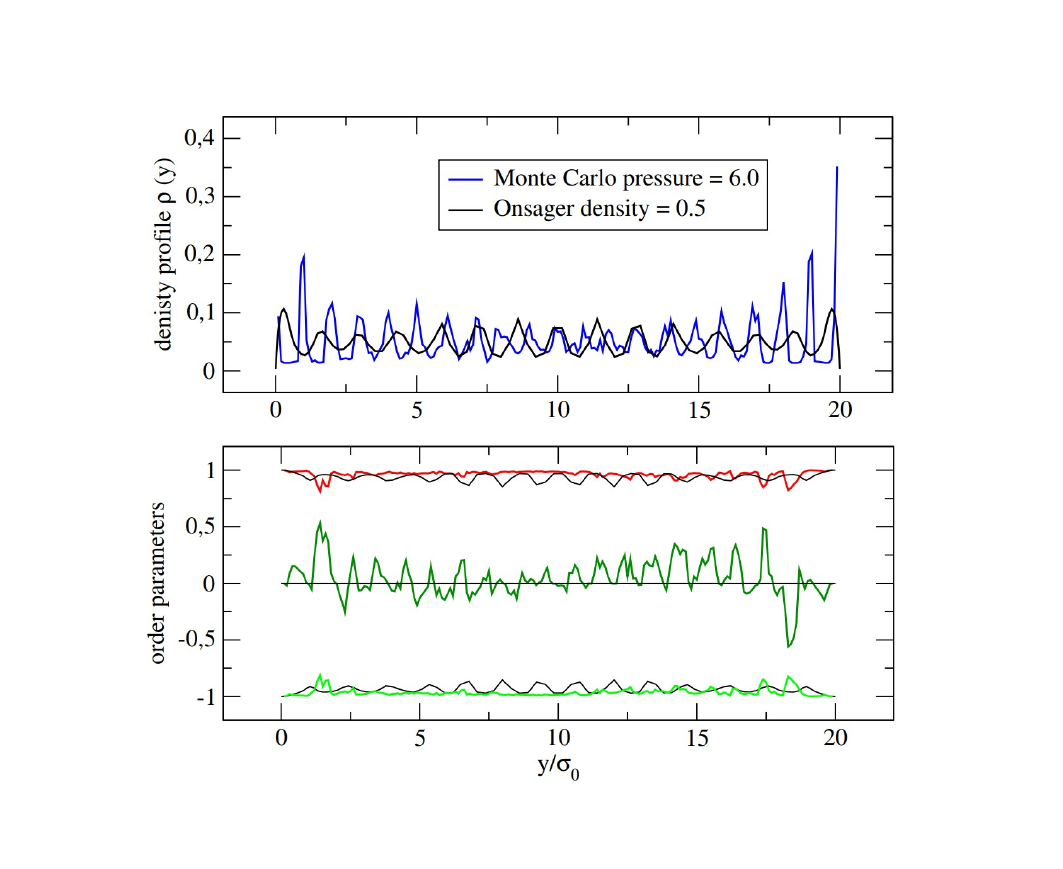}
\end{center}
\caption{Density profile and components of the order parameter tensor for the case of a well ordered planar smectic phase. Above:  a view at the part of  a  2D HGO system with the walls placed at up and down levels.
.
 }
\label{2d_060MC}
\end{figure}

 In Figure (\ref{2d_060MC}) the profiles for a dense system is given. One sees several density
 peaks denoting the structure is layered. The peak in the middle has similar height, yet the other peaks are different. 
Also the number of peaks does not agree. The orientation is parallel to the walls and the orientational profiles in MC and in MD are similar, which is of no surprise, once the system is very ordered and the parameters are close to $1$.

From the above presented results two important conclusions can be drawn. First conclusion is that no bistable orientational arrangement is observed in 2D cases. This indicates that the reason for such an arrangement observed in 3D cases is caused by the surface of ellipsoids, which  are larger in the middle and narrower at the ends,
 allowing for spaces and conditions, where
other ellipsoids tend to place their ends, which finally results in
 their perpendicular to the walls position. It is interesting that such steric origin
 tendency occurs even in the case of dilute systems (see Figure (\ref{fig_elipki})).
 Another important conclusion is drawn while comparing pictures with particles arrangement
  with the density and orientational profiles.
  At the walls the Onsager DFT theory applied without any renormlization  does not reproduce correctly the density profiles.
 This is already seen in dilute systems (Figure (\ref{2d_019MC})).
These wall departures may change even the number of density peaks as shown in the 3D case
(Figure (\ref{fig_plan32})). This effect is rather on the side of the lack of third and higher order terms in the virial expansion of the free energy. For denser systems
one observes however that at the walls we see spatial arrangement (Figure (\ref{2d_022MC})) which requires
inclusion of all space coordinates instead of only one perpendicular to the walls,
but this influence is of smaller importance as compared to the inclusion of higher order terms in virial expansion.

\section{Discussion}

\label{discu}
 Orientational and density profiles of the liquid crystalline system made from HGO particles placed in a slit geometry have been obtained and presented from the Monte Carlo simulations and DFT   theories for 3D and 2D geometry.  Both approaches provide similar results.
One observes, however, some structural departures that increase with the system averaged density. These departures are the result of the second virial approximation in the DFT approach and the lack of  third and higher order terms in the virial expansion of the free energy.
We arrive at a conclusion that these terms play more important role in case of confined materials comparing to the bulk systems. Unfortunately, including such terms in the DFT calculation is not an easy task, mainly due to the computers performance  limitations. To our knowledge, there has been no attempt so far made to tackle this problem in confined systems.
Nevertheless, even on the grounds of the second virial approximation still
a lot work can be done to understand mechanisms governing the physics of confined liquid crystals.
Work and interest of researchers focuses here on the issue of the
interactions with the walls and its influence. It seems, however, that the
most crucial resultant property is the local density at the walls.
It directly emerges from the surface interactions, yet there is no simply way
to predict its value just by changing the  interactions parameters. One must perform
the whole procedure of solving theoretical equations or perform MC simulations.
In principle, we all are aware that density decides about liquid crystalline phase. If there is a strong density peak at the walls
one can expect then  a strong orientational order of molecules. This order influences parts of LC adjacent to direct surfacial regions, and these adjacent parts influences the parts placed next to them and so on. As a result the number of the density (smectic) peaks can be changed (depending on the situation).
In the present paper we show that by changing  penetrability of the molecules (in the DFT theory) we can change (or renormalize) surfacial density and, as a result, it is possible to find a solution that exhibits the correct number of the density peaks and the structural and orientational profiles are very close to the MC results. Examples were shown for the case of nematic, the onset of the smectic and well ordered smectic in 3D case.
One should however be cautious with such a renormalization of the surface density, since upon applying  it the conditions used in the MC and DFT are not exactly the same, (although the results are close), so it is not fully legitimate or universal.
We present the above comparisons rather as an interesting result
that reveals the importance of the local density at the surfaces.

The current paper presentation clearly highlights then the role of the local surfacial density.
 This local density, however, is the result of three different factors:
particles shape, the contact potential function and the averaged density.
 Although we all realize that the contact potential functions are the most important
element that influences confined materials,
there are many  aspects of these functions that may determine this influence.
Besides the strength and anisotropy of  interactions at the confinement regions, the discussed in the present paper  penetrability factor, it is also  the
architecture of the surfaces themselves that is of vital importance.
In the last aspect the most popular idea used in practice is  moderating surfaces with a thin layer of polymer liquid crystal, which, on the other hand,  seems to be  one of more sophisticated theoretical tasks.
It has occurred indeed that
changes of the surface density can be done by introducing the walls made from grafted
polymer chains.
 In \cite{Lange2002JChemPhys,Lange2002Eur,Lange2002Comp}
 for a system made of ellipsoidal Gay-Berne particles
Lange and Schmid  have presented the possibility
of an anchoring transition between tilted and homeotropic arrangements
upon manipulating the grafting density.

Besides the surface density it is the particles shape which determines the orientational behavior. It may also  lead to new effects like presented here eigenvalue exchange problem
or bistable arrangement where particles directly at the walls are placed planarly and particles inside the sample attain perpendicular to the walls orientation.
This was the case for all densities used here and occurred in the MC simulations as well as in the DFT calculations for three-dimensional case. Interestingly, this effect does not occur in 2D.
The conclusion is that it is the surface of the particles that is responsible for such a bistable arrangement.
We suggest that it is the fact that ellipsoids become narrower at the ends which
entails occurrence of the force that reorients the particles. In contrast, in the case of spherocylinders this force may  be absent.
This conclusion can be supported  by the results of
the MC simulations of spherocylinders at the single wall performed by Dijkstra {\it{et al. }} \cite{Marjolein}, where the hard wall promoting planar alignment
induces  a thick layer of planar nematic. The spherocyllinders used in this paper were quite long, hence it is not fully certain that shorter particles will behave in the same manner.

It is very interesting that the bistable ordering  has been also reported in the case of particles interacting through
a continuous potential.
In \cite{Palermo} Palermo {\em{et al.}} have shown by the use of the MC simulations of the Gay-Berne particles placed on the graphite surface, a discontinuous change in anchoring from planar to normal on going from the first
to the second adsorbed layer. This is exactly the same effect as discussed in the present paper. In view of the HGO properties discussed here we can claim then that this effect is
caused by the steric part of the Gay--Berne potential.
Since up to now this potential is the most realistic one, the question arises
about the physical mechanisms which promotes uniform planar (parallel to the walls) alignment in the confined geometries, of slit or cylindrical shape, which are met in reality.
It is an open question now whether the Gay-Berne potential is sufficient to describe the realistic LCs in confinement
or whether the surface potentials (contact functions) must be of much stronger influence to overcome the tendency that leads to simultaneous bistable configurations (like in \cite{Barmes}).

\section{Acknowledgments}
This work was supported by Grant No. DEC-2021/43/B/ST3/03135 of the National Science Centre in Poland.

\end{document}